\documentclass[twocolumn,showpacs,preprintnumbers,amsmath,amssymb]{revtex4}

\usepackage{graphicx}
\usepackage{dcolumn}
\usepackage{bm}

\begin{document}


\title{Two-Particle Microrheology of quasi-2D Viscous Systems}

\author{V.~Prasad}
\author{S. A.~Koehler}
\author{Eric R.~Weeks}%
\affiliation{ Department of Physics, Emory University, Atlanta, GA
30322 }

\date{\today}

\begin{abstract}
We study the spatially correlated motions of colloidal particles in
a quasi-2D system (Human Serum Albumin (HSA) protein molecules at an
air-water interface) for different surface viscosities $\eta_{s}$.
We observe a transition in the behavior of the correlated motion,
from 2-D interface dominated at high $\eta_{s}$ to bulk
fluid-dependent at low $\eta_{s}$. The correlated motions can be
scaled onto a master curve which captures the features of this
transition. This master curve also characterizes the spatial
dependence of the flow field of a viscous interface in response to a
force. The scale factors used for the master curve allow for the
calculation of the surface viscosity $\eta_{s}$ that can be compared
to one-particle measurements.
\end{abstract}

\maketitle


Diffusion in three dimensions has been well understood since 1905,
when two authors showed that the motion of particles suspended in a
fluid is related to the fluid's viscosity
\cite{einstein1,sutherland1}.  This observation has been generalized
in a technique called microrheology, which measures the thermal
motion of tracer particles introduced in a viscoelastic material.
From the motions of the particles, the material dependent properties
can be determined, such as the elastic modulus, $G'(\omega)$, and
the viscous modulus, $G''(\omega)$ \cite{mason1}. This has been
applied to measure the viscoelasticity of bulk materials such as
polymer solutions \cite{dasgupta1}, biomaterials \cite{gardel1} and
hydrogels \cite{breedveld1}.  A closely related question is the
motion of tracer particles in a two-dimensional system such as lipid
molecules at an air-water interface \cite{sickert1} or lipid rafts
in cell membranes \cite{pralle1}. For example, in a purely viscous
2-D system, one might imagine that the diffusive properties are
related to the two-dimensional viscosity, and that by following the
motion of tracer particles one could determine this viscosity.
However, in most cases of practical interest, a strictly
two-dimensional surface is an idealization and in reality the
surface is adjacent to three-dimensional fluid reservoirs.  For
example, recent experiments study diffusion in biological systems
such as cell membranes \cite{pralle1,gambin1} which are surrounding
a 3-D cell and immersed in a 3-D fluid.  This coupling modifies the
behavior of tracer particles and makes interpretation of the results
trickier \cite{saffman1,fischer1,stone1}.

Furthermore, in many cases in 3D, tracers are known to modify the
structure of the medium in their vicinity, leading to erroneous
measurements of rheological quantities \cite{crocker1}.  Another
possibility is that pre-existing inhomogeneities such as pores in an
otherwise rigid material can entrain the tracers, resulting in
measurements that underestimate the bulk viscoelasticity of the
material in question \cite{gardel1}. To overcome these difficulties,
a new method known as two-particle microrheology has been
established \cite{crocker1}, which looks at the cross-correlated
thermal motions of pairs of particles. The correlated motion of two
beads is driven by long-wavelength modes in the system, and is
therefore independent of the local environment of the tracers. While
two-particle microrheology has been applied to 3-D systems
\cite{crocker1,levine2} where the strain field decays as $1/R$, a
formalism for 2-D interfacial systems coupled to viscous bulk fluids
has not been experimentally determined to date. This is largely due
to the non-trivial nature of the flow field created by the thermal
motion of particles at an interface. An understanding of this flow
field is critical towards determining the true microrheological
behavior of an interface \cite{sickert1,pralle1,gambin1}.

In this Letter, we look at the correlated motions of particles
embedded at an air-water interface in the presence of human serum
albumin (HSA) protein molecules. This is done as a function of
particle separation $R$ and lag times $\tau$, for correlated motion
along the line joining the centers of particles, $D_{rr}(R,\tau)$,
and motion perpendicular, $D_{\theta\theta}(R,\tau)$. The interface
is purely viscous, with $D_{rr}$, $D_{\theta\theta}\sim\tau$. The
correlations show a transition as a function of the surface
viscosity $\eta_{s}$; at high $\eta_{s}$ their behavior is 2-D
dominated; at intermediate $\eta_{s}$ they show crossover behavior;
and at low $\eta_{s}$ their behavior is strongly influenced by the
3-D fluid reservoirs ($\sim1/R$ and $1/R^{2}$ respectively). The
correlated motion of the tracer particles for different surface
viscosities can be scaled onto a single master curve, which agrees
with theoretical predictions \cite{levine1}. The surface viscosity
$\eta_{s}$ determined from the scaling parameters of the master
curve agrees well with $\eta_{s}$ from one-particle measurements,
demonstrating the homogeneity of HSA at an interface.

We use aqueous solutions of HSA over a narrow range of bulk
concentrations ($c=0.03-0.045$ mg/ml) to obtain our interface. At
these bulk concentrations, HSA molecules diffuse to the air-water
interface to form a thin monolayer of size $\sim 3$ nm \cite{Lu1},
thereby creating a surface shear viscosity. The surface
concentration of HSA slowly increases over time \cite{chen1}, and so
the surface shear viscosity $\eta_{s}$ can be varied over a wide
range. The viscosity of the bulk solution is negligibly different
from the viscosity of water, and is assumed to be $\eta=1.0$
mPa$\cdot$s for all our experiments, while the viscosity of air is
considered to be negligible. Micron-sized polystyrene
beads(Interfacial Dynamics Corporation, carboxyl-modified, radius
$a=0.9 {\rm \mu m}$) dispersed in an aqueous solution with 20 $\%$
isopropyl alcohol, are spread at the interface with a syringe needle
to act as tracer particles. The particles are imaged by bright-field
microscopy with a 20$\times$ objective, NA=0.45, at a spatial
resolution of 606 nm/pixel and a frame rate of 30 Hz. For each
sample, short movies of 200 frames are recorded with a CCD
camera(Integrated Design Tools, X-Stream Vision XS-3) that has 1240
$\times$ 1024 pixel resolution, with hundreds of particles lying
within the field of view. The positions of the tracers for every
frame are determined by particle tracking. Of potential concern is
the presence of interactions between particles \cite{aveyard1},
whether due to electrostatic forces, capillary forces, or other
origin.  For all experiments, we measure the pair correlation
function $g(R)$ and find no structure for the separations $R$
considered in our results. Also, our experiments are conducted in
the dilute particle limit, and a control experiment for one of the
surface protein concentrations verifies that our results are
unchanged when the particle concentration is varied by a factor of
4. This ensures that long-range particle interactions are not
present and do not affect our conclusions.

From the particle positions, we determine the vector displacements
of the tracers $\Delta
r_{\alpha}(t,\tau)=r_{\alpha}(t+\tau)-r_{\alpha}(t)$ where $t$ is
the absolute time and $\tau$ is the lag time. Care is taken to
eliminate any global drift of the sample from these vector
displacements. We can then calculate the ensemble-averaged
cross-correlated particle motions \cite{crocker1}
\begin{eqnarray}
D_{\alpha\beta}(r,\tau) = \langle \Delta r_{\alpha}^{i}(t,\tau)
\Delta r_{\beta}^{j}(t,\tau) \delta [r-R^{ij}(t)]\rangle_{i\neq j,t}
\end{eqnarray}
where $i$,$j$ are particle indices, $\alpha$ and $\beta$ represent
different co-ordinates and $R^{ij}$ is the distance between
particles $i$ and $j$. In particular, we focus on the diagonal
elements of this tensor product: $D_{rr}$, which measures the
correlated motion along the line joining the centers of particles,
and $D_{\theta\theta}$, which measures the correlated motion
perpendicular to this line (the off-diagonal elements are assumed to
be uncorrelated, and hence 0). In addition, we also calculate the
one-particle MSD, $\langle\Delta r^{2}(\tau)\rangle$, from the self
terms ($i=j$) in the above expression. In 2-D interfacial systems
coupled to a bulk fluid reservoir, this one-particle MSD is related
to the surface viscosity $\eta_{s}$ by a modified Stokes-Einstein
relation \cite{saffman1,fischer1,stone1}
\begin{eqnarray}
\langle\Delta r^{2}(\tau)\rangle= 4 D'_{s} \tau
\end{eqnarray}
where $D'_{s}=D_{s}[{\rm ln}(2\eta_{s}/\eta a)-\gamma_{E}+O(\eta
a/\eta_{s})]$ , with $D_{s}=k_{B}T/4\pi\eta_{s}$ and
$\gamma_{E}=0.577$ being Euler's constant.  This equation, derived
by Saffman for $\eta_{s}/\eta a \gg1$ \cite{saffman1}, has been
modified by Hughes \cite{hughes1} to work in the limit where
$\eta_{s}/\eta a\sim1$. Equation~2 has been used to describe a
variety of homogeneous systems \cite{peters1} and we use the
measured values of $\langle\Delta r^{2}(\tau)\rangle$ to solve it
for the one-particle surface viscosity, $\eta_{s,1p}$ (refer inset
of Fig.~1 for example).

However, as noted above, one-particle measurements may be inaccurate
if the system is heterogeneous, so we perform two-particle
measurements to check this, and also to further probe the spatial
flow field around the particles.
\begin{figure}[htbp]
\includegraphics[scale=0.45]{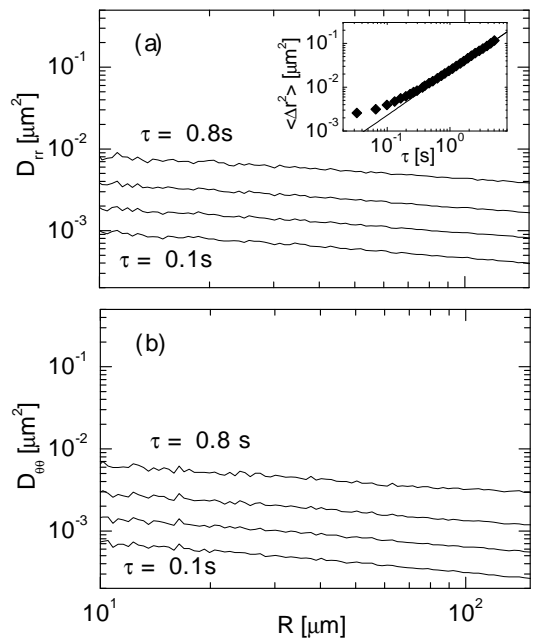}
\caption{\label{fig:1} (a) and (b) Two point correlation functions
$D_{rr}(R,\tau)$ and $D_{\theta\theta}(R,\tau)$ for a sample with
$\eta_{s,1p}=340$ nPa$\cdot$s$\cdot$m, with lag times $\tau = 0.1,
0.2, 0.4$ and $0.8$ s. The evenly spaced correlation functions imply
that $D_{rr}, D_{\theta\theta} \sim \tau$. Inset: One-particle MSD
for the sample (diamonds), where the straight line is a fit giving
$\eta_{s,1p}=340$ nPa$\cdot$s$\cdot$m, using Eqn.~2.}
\end{figure}
In Fig.~1(a), we show $D_{rr}$ as a function of $R$ for different
lag times $\tau$, for a sample with $\eta_{s,1p}=340$
nPa$\cdot$s$\cdot$m. The motion of a tracer particle creates a flow
field that affects the motion of other particles. This flow field
decays as we move further out from the particle; hence the
correlated motions decay as a function of particle separation for
all lag times. From the one-particle measurements, we observe that
HSA monolayers at an interface are entirely viscous at these surface
concentrations, and therefore we also observe that $D_{rr} \sim
\tau$. This is illustrated in Fig.~1(a) where all the $D_{rr}$ are
spaced evenly on a log-plot for lag times $\tau= 0.1, 0.2, 0.4$ and
$0.8$s. A similar linear relationship exists for $D_{\theta\theta}$,
shown in Fig.~1(b). The linear scaling of the correlation functions
enables the estimation of $\tau$-independent quantities $\langle
D_{rr}/\tau\rangle_{\tau}$ and $\langle
D_{\theta\theta}/\tau\rangle_{\tau}$, which depend only on $R$, and
have units of a diffusion coefficient.

Changing the surface shear viscosity $\eta_{s}$, has a substantial
effect on both $D_{rr}$ and $D_{\theta\theta}$. At high $\eta_{s}$,
Fig.~2(a), both $\langle D_{rr}/\tau\rangle$ and $\langle
D_{\theta\theta}/\tau\rangle$ are nearly equal and constant over the
length scales $10< R < 150 \mu$m. At these surface viscosities, the
behavior can be considered to be 2-D dominated, where the bulk fluid
reservoirs have minimal influence. As $\eta_{s}$ is decreased,
Fig.~2(b), we see curvature in these functions and deviation between
their values, over the same range of length scales. At very low
$\eta_{s}$, Fig.~2(c), this deviation is even more pronounced, with
$\langle D_{rr}/\tau\rangle\sim1/R$ and $\langle
D_{\theta\theta}/\tau\rangle\sim1/R^{2}$. At these viscosities, bulk
fluid effects begin to dominate, although the behavior is still 2-D
as the protein molecules are confined to an interface. For all
surface viscosities, the behavior of the correlation functions is
markedly different from what is seen in bulk 3-D systems (where
$D_{rr},D_{\theta\theta}\sim 1/R$), and is sensitively dependent on
$\eta_{s}$.
\begin{figure}[htbp]
\includegraphics[scale=0.52]{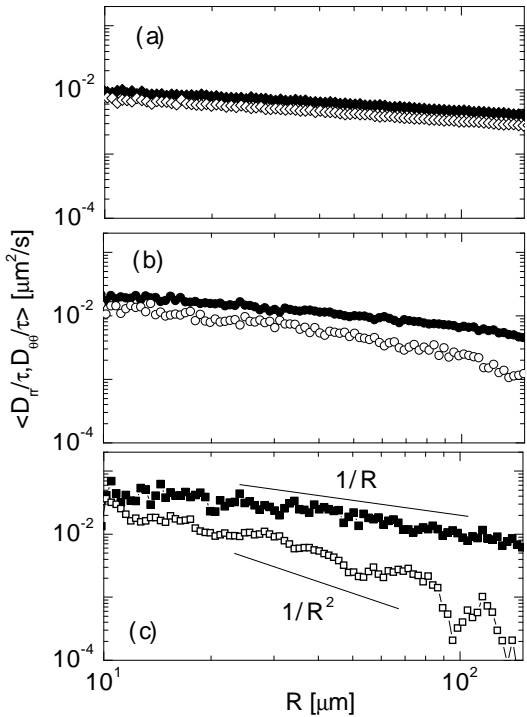}
\caption{\label{fig:2}  Correlation functions $\langle
D_{rr}/\tau\rangle$(solid symbols) and $\langle
D_{\theta\theta}/\tau\rangle$(open symbols) as a function of
particle separation R, for decreasing surface viscosities
$\eta_{s}$. The samples are a) 2-D dominated: $\eta_{s,1p}=340$
nPa$\cdot$s$\cdot$m (diamonds), b) Crossover: $\eta_{s,1p}=72$
nPa$\cdot$s$\cdot$m (circles), c) 3-D dependent: $\eta_{s,1p}=21.3$
nPa$\cdot$s$\cdot$m (squares).}
\end{figure}
Despite the differences in the behavior of $\langle
D_{rr}/\tau\rangle$ and $\langle D_{\theta\theta}/\tau\rangle$ at
low and high surface viscosities, all the data can be scaled onto a
single master curve. We define dimensionless correlation functions $
\bar{D}_{rr,\theta\theta}=\langle
D_{rr,\theta\theta}/\tau\rangle/2D_{s}$ and a reduced separation
$\beta=R/L$, where $L=\eta_{s}/\eta$. Both scale factors $L$ and
$D_{s}$ depend only on $\eta_{s}$; however, we allow them to vary
independently to obtain two independent measures of $\eta_{s}$
($\eta'_{s}$ and $\eta''_{s}$ respectively, in other words, $\beta=R
\eta / \eta'_{s}$, and $D_{s}=k_{B}T/4\pi\eta''_{s}$). Fig.~3 shows
the scaled variables $\bar{D}_{rr}$, $\bar{D}_{\theta\theta}$
plotted against the scaled separation $\beta$. All the data sets
fall on a single master curve, that spans nearly 4 orders of
magnitude. This master curve has the characteristics of the
individual data sets: at small $\beta$ ($\beta\ll1$), the curves are
nearly logarithmic, at intermediate $\beta$ ($\beta\approx 1$), they
show crossover behavior, while at large $\beta$ ($\beta\gg 1$), they
show behavior that asymptotically approaches $1/R$ and $1/R^{2}$ for
$\bar{D}_{rr}$ and $\bar{D}_{\theta\theta}$ respectively. The solid
lines are fits obtained from theoretical calculations of the
response of an interface to an in-plane point force
\cite{levine1,stone1}, and have the form
\begin{eqnarray}
&\bar{D}_{rr}= [\frac{\pi}{\beta}
H_{1}(\beta)-\frac{2}{\beta^{2}}-\frac{\pi}{2}(Y_{0}(\beta)+Y_{2}(\beta))]\nonumber\\
&\bar{D}_{\theta\theta}= [\pi
H_{0}(\beta)-\frac{\pi}{\beta}H_{1}(\beta)+\frac{2}{\beta^{2}}
-\frac{\pi}{2}(Y_{0}(\beta)-Y_{2}(\beta))]
\end{eqnarray}
where the $H_{\nu}$ are Struve functions, and the $Y_{\nu}$ are
Bessel functions of the second kind. More importantly, up to a scale
factor, Eqn.~3, and therefore the master curve, also characterizes
the spatial dependence of the flow field at an interface in response
to a perturbation, such as thermal motion of tracers. To our
knowledge, this is the first experimental mapping of this flow field
over such a wide range of length scales.
\begin{figure}[htbp]
\includegraphics[scale=0.35]{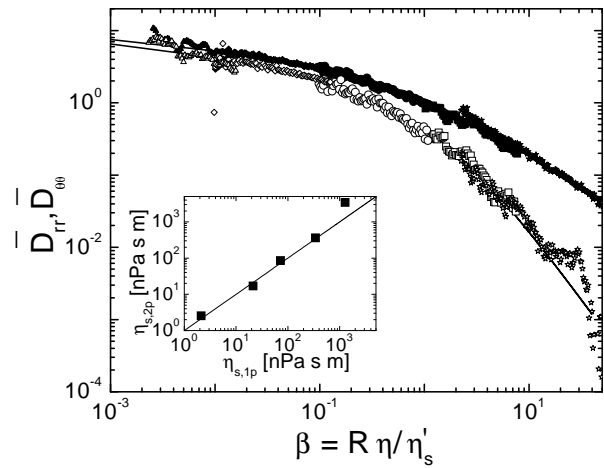}
\caption{\label{fig:3} Master curve of scaled variables
$\bar{D}_{rr}$(solid symbols) and $\bar{D}_{\theta\theta}$(open
symbols) as a function of reduced particle separation $\beta$.
Symbols are same as in Fig.2, with two additional data sets
($\eta_{s,1p}=2.1$ nPa$\cdot$s$\cdot$m; stars and $\eta_{s,1p}=1275$
nPa$\cdot$s$\cdot$m; triangles). Solid lines represent theoretical
fits to the data. Inset: One- and two-particle surface viscosities
for the five samples shown in the master curve. The straight line
has a slope of 1, indicating an equality between the two
viscosities.}
\end{figure}
The inset to Fig.~3 describes the scale factors used to create the
master curve. For all the samples, the two independent estimates of
the two-particle viscosity, $\eta'_{s}$ and $\eta''_{s}$, are within
15$\%$ of each other; we therefore plot their average,
$\eta_{s,2p}$, against the one particle viscosity $\eta_{s,1p}$. For
most of the samples, the one- and two-particle measurements agree
reasonably well with each other. However, at the highest viscosity,
and therefore, the highest surface concentration of HSA, the two
measurements deviate well beyond our experimental error. We believe
this is a consequence of heterogeneity in the system; one
possibility could be the formation of condensed phases at the
interface at such high surface concentrations of HSA. While this
would explain the underestimation of the true surface viscosity by
the one-particle measurement, such an assumption deserves further
study.
\begin{figure}[htbp]
\includegraphics[scale=0.35]{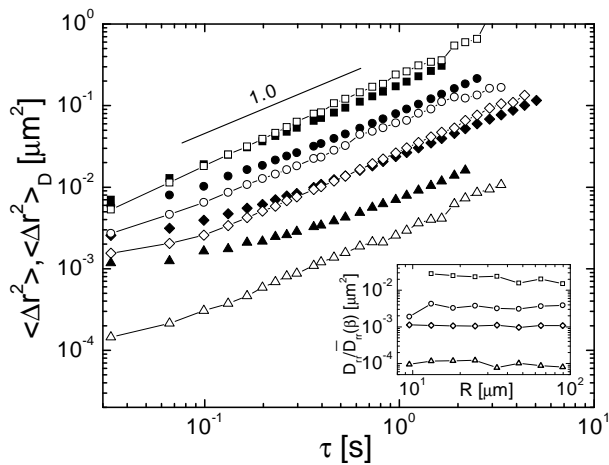}
\caption{\label{fig:4} One-particle (solid symbols) and two-particle
(open symbols) MSDs for four of the five samples shown in Fig.3.
Inset: radial scaling of $D_{rr}$ for the different samples at a lag
time $\tau=0.53$s.}
\end{figure}

Conventional microrheology uses the one-particle MSD, $\langle\Delta
r^{2}(\tau)\rangle$, to determine $G'(\omega)$ and $G''(\omega)$
\cite{mason1}; these quantities are directly compared to bulk
rheological measurements. Since the one-particle measurements are
inherently local in nature, a more accurate approach is to determine
the two-particle MSD \cite{crocker1}, $\langle \Delta r^{2}(\tau)
\rangle_{D}$. Analogous to \cite{crocker1}, we calculate this
quantity for our 2-D viscous systems by extrapolating the long
wavelength fluctuations of the medium to the bead size, which gives
\begin{eqnarray}
\langle \Delta r^{2}(\tau) \rangle_{D}
=2\langle\frac{D_{rr}(R,\tau)}{\bar{D}_{rr}(\beta)}\rangle_{R} [{\rm
ln} (2\eta_{s}/\eta a)-\gamma_{E}+O(\eta a/\eta_{s})]
\end{eqnarray}
This expression can readily be generalized for viscoelastic
interfaces, i.e, with a non-zero $G'(\omega)$; however, such
measurements are beyond the scope of this study. In practice, we
confirm that $D_{rr}(R,\tau)/\bar{D}(\beta)$ is nearly constant over
the length scales studied, $9<R<100{\rm \mu m}$, shown in the inset
to Fig.~4 for a specific lag time ($\tau=0.53$s). The averaged
quantity $\langle D_{rr}(R,\tau)/\bar{D}(\beta)\rangle_{R}$ is then
calculated for all lag times to obtain $\langle \Delta r^{2}(\tau)
\rangle_{D}$. From Fig.~4, we see that for all the samples, $\langle
\Delta r^{2}(\tau) \rangle_{D}$ is purely diffusive, as expected for
a viscous system. At short lag times, the turnover in the
one-particle MSDs, caused by resolution-limited noise, is
significantly reduced in the two-particle MSDs. Finally, at long lag
times, $\langle \Delta r^{2}(\tau) \rangle_{D}$ is equal to $\langle
\Delta r^{2}(\tau) \rangle$ within experimental error, except for
the highest viscosity (as expected, since $\eta_{s,1p}$ deviates
from $\eta_{s,2p}$ for this sample). These observations provide
conclusive evidence of the accuracy of two-particle measurements
over local probes of the rheology.

The verification of two-point microrheological techniques for a
quasi-2D systems has applications for the study of inhomogeneous
materials at an interface. Any significant variation between
$\langle \Delta r^{2}(\tau)\rangle_{D}$ and $\langle \Delta
r^{2}(\tau)\rangle$ indicates the presence of heterogeneities, and
the estimation of rheological quantities from the motion of tracer
particles can be modified to reflect this. Future work will involve
studying lipid molecules at an interface under
compression/expansion, which creates domains such as liquid expanded
or liquid condensed phases. We expect that our two-particle
measurements will provide accurate measurements of the surface
viscosity even with the presence of these heterogeneities. Other
possible areas of future research are biological interfaces such as
cell membranes, with the diffusing entities being protein aggregates
or lipid rafts. However, two-particle microrheology can only probe
heterogeneities of the order of the bead size and above, which may
not always be applicable for lipid systems with molecules in the
nanometer scale. It should also be pointed out that for cases where
$\eta_{s}$ is very small, the correlation functions
($D_{\theta\theta}$ in particular) die out rapidly, making an
estimation of $\eta_{s,2p}$ from the two-particle measurements
difficult. However, one-particle microrheology is equally inaccurate
in this limit, as the bulk viscosity will dominate over surface
effects, making such measurements challenging. Nonetheless, it would
be extremely interesting to test the limits of the scaling behavior
in this regime.

\end{document}